\documentclass{emulateapj}
\usepackage{apjfonts}
\newcommand{\oiii}{[\ion{O}{3}]}

\newcommand{\feii}{\ion{Fe}{2}}

\newcommand{\alphaox}{\ensuremath{\alpha_{\rm ox}}}
\newcommand{\lbol}{L_{\rm bol}}
\newcommand{\ledd}{L_{\rm Edd}}

\newcommand{\chandra}{{\em Chandra}}

\newcommand{\xmm}{{\em XMM-Newton}}
\newcommand{\mbh}{\ensuremath{M_{\rm BH}}}
\newcommand{\eddratio}{\ensuremath{L_{\rm bol}/L_{\rm Edd}}}
\newcommand{\msun}{M_\odot}
\newcommand{\gammahr}{\Gamma_{\rm HR}}
\newcommand{\gammafit}{\Gamma_{\rm fit}}

\def\lax{{$\mathrel{\hbox{\rlap{\hbox{\lower4pt\hbox{$\sim$}}}\hbox{$<$}}}$}}
\def\gax{{$\mathrel{\hbox{\rlap{\hbox{\lower4pt\hbox{$\sim$}}}\hbox{$>$}}}$}}

\shorttitle{X-RAY PROPERTIES OF INTERMEDIATE-MASS BLACK HOLES. III.}
\shortauthors{DONG, GREENE \& HO}

\begin{document}
\title{X-ray Properties of Intermediate-mass Black Holes in Active Galaxies. III. Spectral Energy Distribution and Possible Evidence for Intrinsically 
X-ray--weak AGNs}

\author{Ruobing Dong\altaffilmark{1}, Jenny E. Greene\altaffilmark{1}, and Luis C. Ho\altaffilmark{2}}

\altaffiltext{1}{Department of Astrophysical Sciences, Princeton
University, Princeton, NJ 08544; rdong@princeton.edu,  jgreene@princeton.edu}
\altaffiltext{2}{The Observatories of the Carnegie Institution for Science, 
813 Santa Barbara St., Pasadena, CA 91101; lho@obs.carnegiescience.edu}

\begin{abstract}

  We present a systematic X-ray study, the third in a series, of 49 active 
galactic nuclei
  with intermediate-mass black holes (IMBH; $\sim 10^5-10^6\msun$)
  using \chandra\ observations.
  We detect 42 out of 49 targets with a 0.5--2 keV
  X-ray luminosity $10^{41}-10^{43}$ erg s$^{-1}$. We perform spectral
  fitting for the 10 objects with enough counts ($>$200), and they are all
  well fit by a simple power-law model modified by Galactic
  absorption, with no sign of significant intrinsic absorption. While
  we cannot fit the X-ray spectral slope directly for the rest of the sample, we estimate it from
  the hardness ratio and find a range of photon indices consistent
  with those seen in more luminous and massive objects. The X-ray-to-optical
  spectral slope ($\alphaox$) of our IMBH sample is systematically
  flatter than in active galaxies with more massive black holes, consistent with the
  well-known correlation between $\alphaox$ and UV luminosity.  Thanks
  to the wide dynamic range of our sample, we find evidence that
  \alphaox\ increases with decreasing \mbh\ as expected from accretion disk models, where
  the UV emission systematically decreases as \mbh\ decreases and the
  disk temperature increases.  We also find a long tail toward low 
  \alphaox\ values.  While some of these sources may be obscured,
  given the high \eddratio\ values in the sample, we 
  argue that some may be intrinsically 
  X-ray--weak, perhaps owing to a rare state that radiates very little 
  coronal emission.

\end{abstract}


\section{INTRODUCTION}\label{sec:introduction}

Broad-band spectral energy distributions (SEDs) provide rich
diagnostics on the accretion process in black hole (BH) studies
\citep[e.g.,][]{elv94,ho99,vas09}.  Not only are
SEDs important for measuring the total bolometric luminosity of the
active galactic nucleus (AGN), for constraining models of
accretion disks, and for understanding how BHs
grow, they also determine the impact of a BH on its surroundings.  If
supermassive BHs truly play an important role in the evolution of
galaxies, we must understand how fundamental properties of the AGN,
such as BH mass and accretion rate, impact the SED.

The study of accretion onto stellar-mass BHs is quite mature.
Stellar-mass BHs have accurate BH mass measurements derived from the
orbits of their stellar companions \citep[e.g.,][]{can10} and because the
timescales are quite short, it is possible to watch individual systems
change luminosity states over many orders of magnitude, thus
determining how the SED changes as a function of the mass accretion
rate \citep[e.g.,][]{rem06,fen09,stu11}. Since the accretion disks around stellar-mass 
BHs peak in the X-rays when the BHs are in a high state, there are
also direct measurements of the peak emission from the BH.  In
contrast, with a small number of exceptions, we do not have reliable BH
mass measurements for supermassive BHs in AGNs. Supermassive BH 
accretion disks peak in the far-UV, where we cannot obtain direct 
observations \citep[e.g.,][]{shi78}. Furthermore, changes in
accretion rate occur over prohibitively long timescales in general, so we can 
only probe a range of accretion rates by looking at large populations.  Then, we 
are at the mercy of particular samples, and selection effects substantially 
complicate our efforts to disentangle the effects of $\mbh$ and Eddington ratio on 
the observed SEDs.  

Thanks to the Sloan Digital Sky Survey \citep[SDSS;][]{yor00},
we were able to find a large sample of accreting BHs with the lowest
BH masses known ($10^5-10^6\msun$).  Greene \& Ho (2004; hereinafter GH04)
systematically searched for and found an initial sample of 19 IMBHs from the
first SDSS data release (DR1). With SDSS
DR4, two larger samples have been identified, by Greene \& Ho (2007b;
hereinafter GH07) and \citet{dong12}. In GH07, the BH mass is inferred
from the photoionized broad-line region (BLR) gas 
as $\mbh=fRv^2/G$, where $R$ and $v$ are the radius and
velocity dispersion of the BLR gas, and $f$ is a scaling factor that
accounts for the unknown geometry of the BLR, assumed here to be
spherical ($f=0,75$; \citealt{net90}). The BLR velocity dispersion is
derived from the H$\alpha$ line width, and the BLR radius is inferred
from the AGN luminosity \citep{gre05}, using the so-called
radius-luminosity relation.  The relation between BLR radius and luminosity 
is derived from reverberation mapping of AGNs,
for which radii are measured based on the delay between variations in
the AGN photoionizing continuum and BLR line emission
\citep[e.g.][]{kas05,gre05,ben09}. 

BH masses derived in this way are indirect, and they rely on the
radius-luminosity relation, which is susceptible to large systematic
errors (due to uncertainties in the BLR geometry and kinematics; e.g.,
\citealt{kro01}).  It is certainly true that on a galaxy-by-galaxy
basis we may well be fooled by high-mass outliers. However, we have
some confidence that the average BH masses in our sample are low.
In one case, reverberation mapping points to a very low 
mass for a galaxy from our sample \citep{raf11}.  For the rest, we 
must rely on indirect scalings with the host galaxy properties.
Overall, the host galaxies have low mass.
They are typically about a magnitude below
$L^*$ \citep{bar05,gre08,xia11,jia11a} and the stellar velocity
dispersions are also low \citep{bar05,xia11}. The BH masses expected from
the stellar velocity dispersion measurements are typically
$10^5-10^6~\msun$ for our sample \citep{xia11}.  Based on bulge masses,
instead, the expected BH masses would be an order of magnitude or so
larger \citep{jia11a,jia11b}. 
As further caveats, we note that we have extrapolated both 
the $M-\sigma$ and $M-M_{\rm bulge}$
correlations, and also that the $M-\sigma$ relation may break down
in spirals \citep[e.g.][]{gre10,hu08,kor11}.

We already have some interesting constraints on the broad-band SEDs for the AGNs in the GH07 sample.
We have found that the sources are very radio-quiet on average
\citep{gre06}.  We have seen very indirect evidence that the far-UV
slopes are steeper than in more massive systems \citep{gre05,lud12}.
Finally, we saw tantalizing hints that the sources are relatively
X-ray bright \citep{gre07a,min09,des09}.  In this paper, we present
\chandra\ observations for a larger sample of sources from GH07.  The
current sample contain 49 sources drawn from the 174 presented in
GH07, excluding the less secure $c$ sample (GH07 flagged a subset of their detections
as low-significance in cases that they could not be confident in the
presence of a broad H$\alpha$ line from the SDSS spectrum alone). The general selection
strategy is to observe the nearest galaxies, with the average redshift of
the 49 objects being $\sim0.05$ comparing to $\sim0.09$ for the 174 GH07
objects. The observations are 2~ks ``snapshots,'' which allow us to
observe a large number of targets within a limited time.  We will
argue that our X-ray observations point to real changes in the
structures of accretion disks as a function of $\mbh$.

The structure of this paper is as follows. We present the data
analysis and spectral fitting processes in Section~\ref{sec:data},
where we try various models to fit the spectra. Then in Section~\ref{sec:xray} we discuss the X-ray properties of the sample. The broader SEDs of the sample is discussed in Section~\ref{sec:alphaox}, specifically we focus on the ratio of optical/UV to X-rays. We identify a potential subgroup of
intrinsically X-ray--weak candidates in Section~\ref{sec:xrayweak}, and discuss the possible origins of their X-ray weakness. A short
summary is provided in Section~\ref{sec:summary}. The cosmological parameters
assumed in this paper are $H_0 = 71$ km s$^{-1}$ Mpc$^{-1}$, $
\Omega_m = 0.27$, and $\Omega_\Lambda = 0.75$ \citep{spe03}.

\section{Observations and Data Analysis}\label{sec:data}

The \chandra\ \citep{wei96} observations in this work were taken between 2009 
September and 2010 September (proposal number 11700259). They are snapshot 
observations, with effective exposure times of $\sim$2 ks.  As in Greene \& 
Ho (2007a; Paper I) and Desroches et al. (2009; Paper II),
the Advanced CCD Imaging Spectrometer (ACIS; Garmire et al. 2003) was 
used, and images were obtained at the aim point of the S3 CCD in faint mode. 
We only read out 1/8 of the chip to reduce the effects of pile-up.

The basic data reduction steps are described in Papers~I and II.  Here we give 
only a brief summary.  We begin with the standard Level 2 event files 
processed by the \chandra\ X-ray Center, which already have been corrected for
cosmic rays filtered for good time intervals. Events below 0.3 keV and 
above 8 keV are rejected in the analysis to avoid calibration uncertainties at 
low energies and to limit background contamination at high energies 
\citep{gal08}.  We follow the standard
procedure of detecting faint point sources in \chandra\ observations,
and we use the task {\tt celldetect} in the \chandra\ Interactive
Analysis of Observations (CIAO) package\footnote{\tt http://cxc.harvard.edu/ciao/download/doc/detect\_
manual/cell\_theory.html}
with default parameters to detect 
sources and extract their centroid positions. We set an initial detection threshold of
signal-to-noise ratio (S/N) $\geq3$. In every case {\tt celldetect} 
yields 0 or 1 detection, with a detection being a point source at the aim 
point. We detected 42 out of 49 sources.  All but two were detected with a 
signal-to-noise ratio (S/N) $\geq 3$, while the final two had $3>$S/N$\geq 2$.
A detection threshold
of S/N$\geq3$ is the standard value in \chandra\ faint source observations,
as recommended in CIAO. However, we note that in the regime of nearly
zero background, a low-S/N detection such as S/N=2 can still be a
statistically significant detection. Studies of simulated \chandra\ data\footnote{\tt http://cxc.harvard.edu/ciao/download/doc/detect\_
manual/cell\_false.html}
show that the false detection rate with S/N$\geq1$ in 10 ks ACIS on-axis
observations is about 0.1 per four imaging chips in ACIS-I. The
false detection rate of S/N$\geq2$ in our 2 ks observations should be even
lower.
Interestingly, we note that the detection rate for this sample with 2~ks 
snapshots (42/49) is actually higher than that in the GH04 sample with 5~ks
observations (13/18; Paper~II).  This apparent peculiarity is presumably due to 
the smaller distances of the current sample, whose average redshift is 
$\sim0.05$, compared to $\sim0.09$ for the GH04 sample. The position 
difference between \chandra\ and SDSS is less than 1\arcsec\ for all sources, 
and less than 0\farcs5 for $90\%$ of the sample, which is the spatial 
resolution of \chandra.

The on-axis point-spread function of \chandra\ 
contains $95\%$ of the encircled energy within $1\arcsec$. We therefore 
extract counts from a $2\arcsec$-radius circle centered on the source, in the 
soft (0.5--2 keV; $C_s$) and hard (2--8 keV; $C_h$) bands.  We measure 
background rates from a concentric annulus with inner radius 7\arcsec\ and 
outer radius 15\arcsec, and background-corrected count rates are calculated 
using the CIAO task {\tt dmextract}. The background rates within the aperture 
are always very low ($<1$ count~s$^{-1}$ for all cases). For the undetected 
objects, we calculate an upper limit of the counts in the full band necessary 
to make a theoretical detection with a S/N of 2 using our source detection 
procedure.  We follow the CIAO help page\footnote{\tt 
http://cxc.harvard.edu/ciao/download/doc/detect\_
manual/cell\_theory.html} 
on how to determine the S/N given the source and background counts, 
then divide this full-band count upper limit into $C_s$ and $C_h$, assuming 
a photon spectral index $\Gamma=2$ [$N(E)\propto E^{-\Gamma}$], which is the 
average value from Papers~I and II and from the current sample (see below). We 
note here that this procedure is different from that used in Paper~II, 
where the $C_s$ and $C_h$ upper limits for non-detections were calculated to 
have a theoretical detection with a S/N of 2 in each band separately, given the 
background counts in each band. The current strategy is more accurate in 
the sense that it is a direct analog to the detected cases, while the previous
strategy yielded an unrealistically high upper limit. When they are used 
in the analysis below, the undetected objects from Paper~II are reprocessed 
following the new strategy.

\subsection{Hardness Ratios}

As in \citet{gal05} and in Papers~I and II, we measure the hardness ratio, HR 
$\equiv (C_h-C_s)/(C_h+C_s)$, for all the detected targets.  This gives a 
crude estimate of the spectral shape, from which we can then infer a spectral 
index $\gammahr$. We use the CIAO task {\tt psextract}\footnote{Since we use 
$-120^o$ ACIS data taken in FAINT mode, which have gone through Reprocessing 
III, we run {\tt mkacisrmf} after {\tt psextract} as recommended by the CIAO 
team (http://cxc.harvard.edu/ciao/why/mkacisrmf.html), and we rerun 
{\tt mkarf} after that to match the energy grids for the ARF and RMF files.} 
to get the instrumental response functions, both the auxiliary response file 
(ARF) and the redistribution matrix file (RMF). We use the spectral-fitting 
package XSPEC \citep{arn96} to generate artificial spectra with known spectral 
slopes ($0<\Gamma<4$) and Galactic absorption (Dickey \& Lockman 1990; 
Table 1).  Then we ``observe'' the artificial spectra using the ARF and RMF 
extracted for each observation, and measure the HR for each input $\Gamma$. By 
comparing each observed HR with the HR from artificial spectra, we infer 
$\gammahr$, as listed in Table~\ref{tab:x-rayproperty}. Below we check the 
validity of this method by comparing $\gammahr$ with $\gammafit$ from actual 
spectral fitting of bright sources, and the comparison shows that the HR-based 
spectral indices are reasonable (but see the caveat in Section~\ref{sec:xrayspectra}). For the sake of uniformity, all the analysis 
in this paper, unless specified otherwise, is done using $\gammahr$.

\subsection{Spectral Fitting}\label{sec:spectral}

We perform spectral fitting for the 10 sources with sufficient total counts 
($>200$ in the entire band).  We bin the event files to achieve at least 20 
counts per bin to ensure the validity of $\chi^2$ statistics. We use the task 
{\tt psextract} to extract the spectra, using the same aperture and background 
region as above, and the appropriate RMF and ARF files.  For most sources we 
limit the energy range to $\sim 0.3-5$ keV to avoid large uncertainties due to 
detector response and low counts, while for some of the faintest ones we 
reduce this range even further (see Table~\ref{tab:x-rayproperty}).

\begin{figure*}
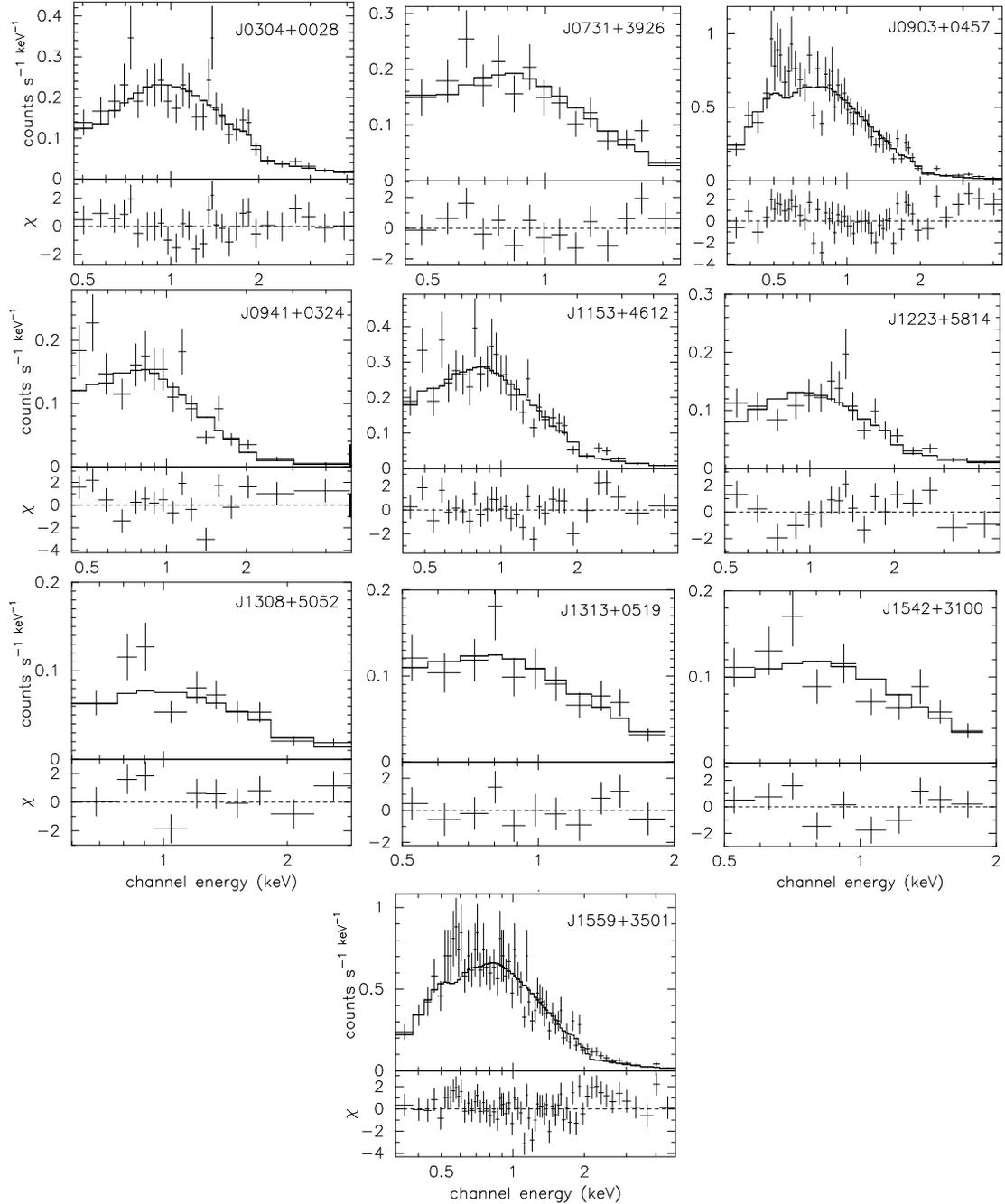

\epsscale{0.33}
\plotone{11441} \epsscale{0.30} \plotone{11442} \plotone{11448}
\epsscale{0.33}
\plotone{11451} \epsscale{0.30} \plotone{11462} \plotone{11464}
\epsscale{0.33}
\plotone{11467} \epsscale{0.30} \plotone{11468} \plotone{11477}
\epsscale{0.33}
\plotone{11479}
\figcaption{Spectral fits for the 10 sources with $\geq200$ counts. Each source
is fitted with a single PL modified only by Galactic absorption. The bottom 
panel of each plot shows the residuals of the fit, expressed in terms of 
standard deviation $\chi$ with error bars of 1.  Note that the energy range is 
adjusted in each individual plot due to different total source counts.
\label{fig:spectrum}}
\end{figure*}

All of the spectra are fitted with XSPEC with a single absorbed power-law (PL)
model, with absorption fixed to the Galactic value \citep{dic90}. We list the 
fitted parameters as well as the $\chi^2$ statistics in Table~\ref{tab:fit}
and provide the spectra and their fits in Figure~\ref{fig:spectrum}. The 
quoted errors are at the $90\%$ confidence level. With the reduced 
$\chi^2_\nu$ being consistent with 1 and the fitting residuals randomly 
distributed, a simple absorbed PL model proves to be adequate for all the
sources, with no additional components needed. To check if there is any sign 
of intrinsic absorption, we fit the spectrum with a PL model modified by 
an absorption component in addition to the fixed Galactic component. The 
results are generally worse than the simple Galactic absorbed PL model, as 
indicated by their $\chi^2_\nu$ being farther from 1.  There is no evidence for 
intrinsic absorption in any of the sources, at the $90\%$ confidence level;
this is consistent with the conclusions from Papers~I and II.  We note that 
this result might be influenced by the low source counts and limited spectral 
coverage of our snapshot observations.

We also try to look for possible evidence for a soft excess in the spectra. 
Soft excesses above a fiducial PL are widely observed in luminous AGNs.  It is 
modeled in various ways, such as a broken PL, a multicolor disk blackbody 
(diskbb), or a Comptonized multicolor disk \citep{wan04, min09}, although its 
origin is still hotly debated (Crummy et al. 2006; Done \& Nayakshin 2007; 
Miniutti et al. 2009).  We fit the spectra using a Galactic absorbed PL+diskbb 
model to see if there is any evidence for a soft excess component in our 
sample.  In general, this model does not provide an obviously better fit to 
the spectra. Most of the values for the fitted inner disk temperature are 
unnaturally high, with $T_{\rm in} \approx 1.5$ keV), to be compared with 
typical values of  $T_{\rm in} \approx 0.15$ keV seen in narrow-line Seyfert 1 (NLS1) galaxies 
\citep{lei99} and in the GH04 objects bright enough to be studied with \xmm\ 
\citep{min09}.  

\begin{figure}
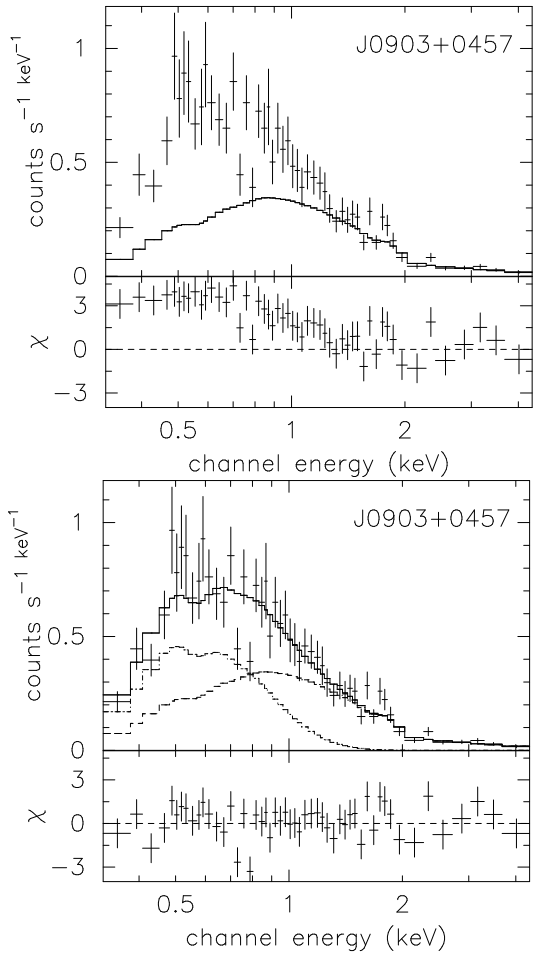

\begin{center}
\epsscale{0.95} \plotone{11448-pl} \plotone{11448-diskbb}
\end{center}
\figcaption{Multi-component fit of J0903+0457. ({\it Left}) A PL fit
restricted to the energy range 1.5--5 keV; when extrapolated toward lower
energies, a soft excess is clearly visible. ({\it Right}) A model consisting 
of the previous PL plus an additional {\tt diskbb} component (with the solid 
curve showing the combination and the dashed curves showing the two 
individually) provides a reasonable fit to the entire spectrum.  The bottom 
panel of each plot shows the residuals of the fit, expressed in terms of 
standard deviation $\chi$ with error bars of 1.
\label{fig:J0903+0457_multi}}
\end{figure}

Another way to look for the soft excess is to fit the hard-band spectrum using
a single absorbed PL, and then extrapolate it into the soft band to see if 
there is any additional excess above the PL (Paper~I; Miniutti et al. 2009). 
Unfortunately, we can only do this for two sources, SDSS~J0903+0457 and 
SDSS~J1559+3501, which have enough counts in the hard band to ensure a 
reliable fit. Following Paper~I, we fit the data using only photons with energies
above 1.5 keV with a PL 
component modified by fixed Galactic absorption.  In the case of 
SDSS~J1559+3501, there is no excess when the PL is 
extrapolated to the soft band. In fact, the fitted parameters agree, within the
errors, with the fit to the full-band spectrum (0.3--5 keV), indicating that 
the entire spectrum is reasonably described by a single absorbed PL model;  
adding an extra diskbb component is not necessary. However, in the case of SDSS~J0903+0457, as shown in the left panel of Figure~\ref{fig:J0903+0457_multi}, a soft excess clearly stands
out when we fix the power-law component using only photons with
energies above 1.5 keV. The hard-band $\Gamma_h=1.99\pm0.20$ is
significantly flatter than that obtained from fitting the full
spectrum, $\gammafit=2.42\pm0.06$. Adding a diskbb component in
addition to the PL to model the full spectrum (0.3--5 keV) gives a
significant improvement to the fit (right panel of
Figure~\ref{fig:J0903+0457_multi}), resulting in $\chi^2_\nu=61.7/50$
(the power-law component is fixed from fitting photons with energies
above 1.5 keV) over the previous value of $\chi^2_\nu=83.3/50$. The
inner disk temperature of $T_{\rm in}=0.15\pm0.006$ keV is consistent
with values seen in other IMBHs \citep{min09, tho09}.  We conclude
that for this object, a two component model (pl+diskbb) is consistent
with the data, while there is no clear evidence for a soft excess in
the rest of the sample.

However, an 
additional diskbb component has been firmly detected by \citet{min09} for at 
least three out of four IMBH AGNs selected from the GH04 sample, and Thornton 
et al. (2009) show that POX 52 also contains a prominent soft component.
Moreover, using the same technique, \citet{por04} found that soft excesses are 
very common in their sample of higher mass ($10^7\msun<\mbh<10^9\msun$) but 
also high-Eddington ratio ($\eddratio=\lbol/\ledd \approx 0.1-1$) AGNs.   Thus, we again
caution that the non-detection of a soft component in most of our objects may be
an artifact of the low source counts and narrow energy coverage of our 
snapshot observations.  Indeed, none of the objects for which Miniutti 
et al. (2009) detected a soft excess in \xmm\ observations show evidence for
such a component in their shallower \chandra\ observations.

\begin{figure}
\begin{center}
\epsscale{0.95} \plotone{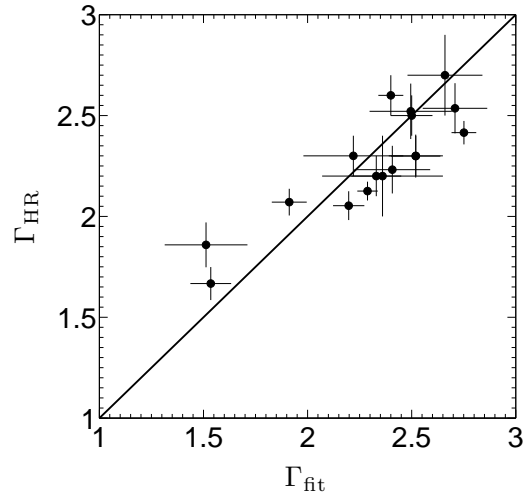}
\end{center}
\figcaption{Comparison of the X-ray photon index $\gammahr$, calculated from
the HR method, and $\gammafit$, which is derived directly from spectral
fitting, for the 17 sources from the combined sample that have enough
counts to allow spectral fitting. The solid diagonal line is the
1:1 relation.  The two photon indices generally agree well within the range of $\Gamma$ probed.
\label{fig:fit}}
\end{figure}

The ``HR comparison" method for estimating the X-ray spectral index $\gammahr$, 
when spectral fitting is not applicable due to low source counts, has been 
used in the literature before (e.g., Gallagher et al. 2005; Papers~I and II) 
and is potentially a very useful method in observations aiming at distant,
faint sources.  Here we verify the reliability of this method, by 
combining data from Papers~I and II and the current work when both spectral 
fitting and HR comparison are available.  Figure~\ref{fig:fit} shows the 
result, where error bars indicate the $90\%$ confidence level for the spectral 
fitting $\gammafit$ and the uncertainties propagated from HR to $\gammahr$. 
The two $\Gamma$ measurements generally agree well, especially at the steep 
end, with an average difference of 0.04 and a standard deviation of 0.18.

\section{X-ray Properties}\label{sec:xray}

Compared with most AGNs previous studied, our optically selected
sample\footnote{Unless otherwise specified, the discussion below
  refers to the observations reported in this paper plus those from
  GH04 (published in Papers~I and II) that remain in the GH07 sample.
  A few of the original GH04 objects no longer satisfy the mass cut of
  GH07, according to the updated mass estimator in GH07.  The final
  combined sample contains 61 sources, 49 from this study and 12 from
  Papers~I and II.  In addition to the revised X-ray upper limits
  described in Section~2, all the optical parameters used in the
  present analysis have also been updated with the most recent values
  from GH07.} is distinguished by two unique features: low BH mass ($\mbh \approx 10^5-10^{6.5}\, \msun$) and high Eddington ratios ($\eddratio 
\approx 1$)\footnote{Although the bolometric correction 
remains uncertain, see below.}.  A major
motivation of our analysis is to investigate whether these two
characteristics impart any noticeable differences on the X-ray
properties of AGNs. The X-ray observations presented here provide 
three basic tools to study the changes in accretion disk and corona 
with mass and Eddington ratio.  First, we look at the X-ray spectrum 
itself.  Second, we extract the X-ray variability properties with time. 
Third, the SED, and specifically the relative X-ray 
to UV flux, provides an important diagnostic of the accretion disk. As 
we will show, our measurements of the first two are not extensive enough 
to provide much new information. Therefore we focus mostly on the SED.
First, we introduce the comparison samples that we use.

\subsection{Comparison with Other Samples and Theoretical Models}
\label{sec:alphaoxcom}

To take full advantage of the dynamic range in BH mass and Eddington
ratio afforded by our sample, we must utilize comparison samples from
the literature.  We compare our AGN sample with two other samples from
recent literature: the ``small clean'' sample of 
(Wu et al. 2012a; hereafter W12), and the sample from Jin et al. (2012; hereafter J12).  These
two samples supplement ours in the parameters space of $\eddratio$ and
$\mbh$. The W12 sample is composed of optically selected quasars from
SDSS DR5, with simultaneous observations in the UV/optical and X-ray
bands using {\it Swift}.  These observations probe wavelengths closer
to the big blue bump (BBB) than ground-based optical observations and
simultaneously measure hard X-rays.  Following W12, we limit the
exposure time to be longer than 10 ks, resulting in a high X-ray
detection rate of $\sim85\%$. Based on the SED fitting, the authors
determined the overall shape of the SED and subsequently derived the
X-ray spectral slope, the bolometric luminosity $\lbol$, and
$\eddratio$. $\mbh$ in this sample ranges from $\sim10^8$ to $10^9\,M_\odot$,
and ${\eddratio}$ ranges from $10^{-1.5}$ to $10^{0.5}$.

The J12 sample contains 51 unobscured type 1 AGNs, with $\mbh$
spanning $\sim 1.5$ dex around $10^8\,M_\odot$, and $\log \eddratio$ about 2
dex centered at 0.3. These objects are X-ray/optically selected to
have high-quality \emph{XMM-Newton} and SDSS spectra in the
literature. The sample is characterized by low reddening in the
optical and low gas columns as implied by their X-ray spectra, so that the
observed properties are likely to be intrinsic. The authors fit the
data with an energetically self-consistent accretion disk model recently
proposed by \citet{don12} to determine the overall shape of the SEDs
and various properties of the sources. The \citet{don12} model
contains three distinct spectral components, all powered
by the energy released by a single accretion flow of constant mass
accretion rate $\dot{M}$, onto a BH with mass
$\mbh$. The model assumes that the disk emission from 
outside a coronal radius $R_{\rm cor}$ thermalizes to a
blackbody, with a color temperature correction. At radii smaller than
$R_{\rm cor}$, the energy released from accretion is split between 
optically thick Comptonized disk emission, which forms the soft X-ray
excess, and optically thin coronal emission above the disk, which
forms the high-energy X-ray tail.

In addition to these two samples, we will also compare our 
targets with other NLS1s. The objects in our sample, with relatively narrow
broad lines ($\rm FWHM_{H\alpha}<2000 \, km\, s^{-1}$), technically
all qualify as NLS1s according to the line width criterion commonly
used to define them, although given their mass cut our objects tend to
occupy the low-mass end of the NLS1 class.  The two groups share a
number of other properties, including the propensity to have high
Eddington ratios (GH07), weak radio emission (Greene et al. 2006), and
low-mass and low-luminosity host galaxies (Greene et al.  2008; Jiang
et al. 2011b), although some optical spectroscopic properties, such as
the distribution of the strength of the \feii\ and \oiii\ lines, are
somewhat different (GH04; GH07).

\subsection{X-ray Spectra}
\label{sec:xrayspectra}

The X-ray spectral index $\Gamma$ is an important parameter that
characterizes the disk-corona structure. Many papers have reported a
correlation between $\Gamma$ (usually measured at $2-10$ keV) and the
Eddington ratio \citep[e.g.,][]{lao97,she08,ris09}.  Theoretically,
the interpretation is that as the UV flux from the disk increases,
Compton cooling is more effective, leading to a steeper power-law slope
\citep[e.g.,][]{pou95,cao09}. On the other hand, more recent studies
of local active galaxies with lower luminosity and lower BH mass have
not confirmed these relations, but rather observe a wide range of
$\Gamma$ values \citep{ai11,kam12}. At lower $\eddratio$, there is a suggestion that the X-ray spectral
slope changes \citep{con09} in analogy with that the spectral state
changes from hard to soft seen when X-ray binaries go from a low to a
high state \citep[e.g.][]{mei01,mac03}. We would like to investigate the
distribution of $\Gamma$ values in our sample, but unlike the aforementioned papers, we
have not measured $\Gamma$ directly.  Instead, we have inferred it
from the hardness ratio based on a limited spectral range.

\begin{figure}
\begin{center}
\epsscale{0.95} \plotone{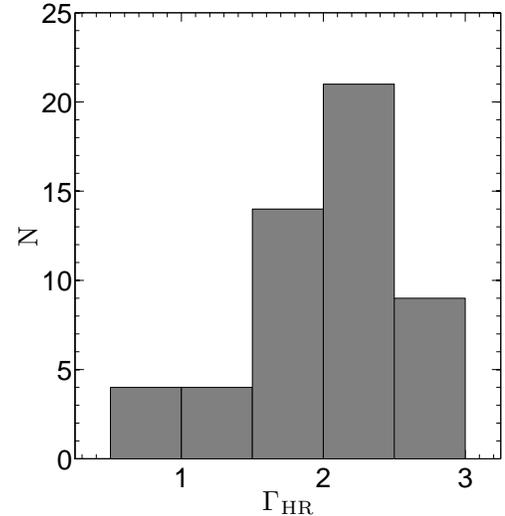}
\end{center}
\figcaption{Distribution of $\gammahr$ for all {\it Chandra}-detected sources.
Each bin is 0.5 in width in $\gammahr$.
\label{fig:histogamma}}
\end{figure}

We use disk models to test the correspondence between our measured
$\Gamma$ values and the intrinsic spectral shape.  We generate mock
AGN spectra using the \citet{don12} model described above (model {\rm
  optxagnf} in XSPEC), which contains a hard power-law component
characterized by an input $\Gamma$. We then ``observe'' these models
using the same instrument response files as our targets, and recover
$\Gamma_{\rm HR}$.  We ask how $\Gamma_{\rm HR}$ compares with
the intrinsic $\Gamma$. Unfortunately, we find that $\Gamma_{\rm HR}$
is not an accurate tracer of $\Gamma$.  Specifically, for values of
$\Gamma > 2$, $\Gamma_{\rm HR}$ is highly correlated with $\Gamma$.
However, when $\Gamma < 2$ (flat), $\Gamma_{\rm HR}$ becomes steep
again because of the soft X-ray excess.  Our sensitivity to hard
photons is too poor to return a high-fidelity measurement of $\Gamma$
in the presence of complex, multi-component spectra. Therefore, we
cannot directly compare our measurements with those based on
\emph{XMM-Newton} spectra that extend to much harder energies.  We note that 
 a similar effect may be present in higher redshift sources as
well \citep[e.g.,][]{wu12a}.  In this work we report our measurements of
$\Gamma$, with the histogram shown in Figure~\ref{fig:histogamma}, but
avoid detailed comparison between our sample and others on 
$\Gamma$-related properties. $\gammahr$ in our sample ranges from $\sim0.5-3$,
with a mean value of 2 and $1\sigma$ deviation of 0.5.

Larger samples of NLS1s with deeper spectroscopy often exhibit 
complex features in their soft X-ray spectra.  These include
absorption edges due to \ion{O}{7} (0.74 keV) and \ion{O}{8} (0.97 keV) and 
Fe K shell emission \citep{bra94,lei97,fio98,tur98,nic99}, although their 
origin is not well understood \citep{ros05}.  While we do not have enough 
sensitivity to detect any specific spectral features in this energy band, we 
do find, in agreement with Paper~I, that most of our spectra show some 
evidence for spectral complexity around 1 keV.

\section{The Ratio of Optical/UV to X-rays}\label{sec:alphaox}

We now consider the broader SEDs of
the sample, and specifically the balance of energy coming out in the
optical/UV emerging from the accretion disk as compared with the X-ray
luminosity from the corona.  We use the ratio of the optical-to-X-ray
flux, $\alphaox$, first introduced by \citet{tan79}.  In general,
there are three ways to interpret the range in $\alphaox$:

\begin{enumerate}

\item In a standard disk, the balance of UV to X-ray photons is
  established via inverse-Compton scattering of UV photons into a
  corona \citep[e.g.,][]{lia77,haa93}.  Thus, changing the temperature
  of the disk, either by changing the BH mass or by changing the
  accretion rate, will lead to changes in the fraction of the disk
  energy that is reprocessed into the corona. Self-consistent magnetohydrodynamics
  simulations describing the fraction of energy emerging in the corona
  for stellar-mass BHs are in development now \citep[e.g.,][]{sch12}.

\item Another way to change $\alphaox$ is to change the structure of the disk
entirely. If at high accretion rates, for instance, the disk
structure changes from a standard $\alpha$-disk to a slim disk
\citep{abr87,abr88}, then the disk may become very small and hot,
and the fraction of energy in the disk may increase \citep[e.g.,][]{min00}.  
Alternatively, the corona may disappear, leading to
high $\dot{M}$ but intrinsically X-ray--weak sources \citep{lei07a,lei07b,wu12b}.
In X-ray binaries, the power-law emission often almost completely 
disappears in sources radiating above a few percent of Eddington \citep[e.g.][]{rem06}.

\item A third way to change $\alphaox$ is through absorption.
  Reddening will lower the UV luminosity and flatten the UV slope,
  while absorption in the X-rays will harden $\Gamma$ and suppress the
  soft X-ray flux. However, as shown by, for example, Vasudevan et al. (2009),
  there is not always a direct connection between red quasars in the optical
  and X-ray--obscured systems.  Thus, variations in absorption with
  time will add noise to intrinsic trends in \alphaox\ \citep{gal06},
  while trends between \alphaox\ and UV luminosity may be
    partially attributable to uncorrected reddening \citep{vas09}.
  Also, we note that different gas-to-dust ratios can cause different amounts
of absorption between X-rays and optical, as has been explored in the 
context of gamma-ray bursts \citep[e.g.][]{per09,zaf11}.

\end{enumerate}

Many papers have studied whether $\alphaox$ changes with either $\mbh$
or $\eddratio$. In this Section, by extending the dynamic range in
$\mbh$ by an order of magnitude, we investigate the statistics of
$\alphaox$ of our sample, and the correlation between $\alphaox$ and
various quantities. Ultimately, we seek differences in the intrinsic SEDs of 
these active nuclei due to their mass or Eddington ratio, and what 
they tell us about disk structure.

\subsection{$\alphaox$ and Correlations with Fundamental 
Parameters}\label{sec:alphaoxstat}

We tabulate $\alphaox$ for our sample in
Table~\ref{tab:x-rayproperty}.  To be consistent with most authors in
the current literature, we adopt the definition
\begin{equation}
\alphaox\equiv-0.3838\log{f_{\rm 2500}/f_{\rm 2~keV}},
\label{eq:alphaox}
\end{equation} 

\noindent
where $f_\nu \propto\nu^{\alphaox}$ is the specific flux. Note that this definition differs
from the original one in \citet{tan79} by a minus sign. Following
Paper~II, we use H$\alpha$ measurements from GH07 to determine the AGN luminosity at 5100 \AA\ 
\citep[$L_{\rm 5100}$,][]{gre05},
\begin{equation}
L_{\rm H\alpha}=5.25\times10^{42}\left(\frac{L_{5100}}{10^{44}\ {\rm erg s^{-1}}}\right)^{1.157}{\rm erg s^{-1}},
\label{eq:alphaox}
\end{equation}
and then assume that the optical continuum follows the
PL $f_\nu\propto \nu^{-0.44}$ \citep{van01, gre07a} to calculate the
monochromatic flux at 2500 \AA\ ($f_{\rm 2500}$). We do not use
the directly measured monochromatic flux at 5100 \AA\ to avoid
potential contamination by the galaxy starlight. Thus we are less
susceptible to reddening than in the case of a direct measurement but
depend on an assumed UV spectral slope. In cases where UV fluxes are
available in the literature (see next paragraph) we find good
consistency with our measurements, suggesting that our assumptions are
valid.  The X-ray monochromatic flux at 2~keV ($f_{\rm 2~keV}$) is
calculated from $\gammahr$ and $f_s$. We note that except for SDSS~J0903+0457, 
which shows a prominent soft
blackbody component in addition to the PL component, the monochromatic
flux at 2~keV comes almost completely from the PL component.


Six objects from the GH04 sample were observed with \xmm\ by 
Dewangan et al. (2008; see also Miniutti et al. 2009).  \citet{dew08} took 
advantage of the Optical Monitor onboard \xmm\ to record simultaneous 
optical/UV and X-ray fluxes, which enabled them to obtain more reliable 
measurements of $\alphaox$ for these objects. From comparison of $\alphaox$ 
between Paper~II and \citet{dew08}, we find a systematic offset of $\sim0.1$, 
with the \chandra\ values being slightly flatter (more X-ray luminous). We 
compare the values of $L_{2500}$ from direct measurements using \xmm\ 
\citep{min09} with our values derived from $L_{\rm H\alpha}$ and find that the 
two agree well (although extinction may still be a factor in individual cases; 
Vasudevan et al. 2010).  On the other hand, our \chandra\ X-ray luminosities 
are systematically higher than those derived from \xmm, which accounts for the 
difference in $\alphaox$. The level of difference, however, is small, and 
does not affect any of the conclusions below.

\begin{figure}
\begin{center}
\epsscale{0.95} \plotone{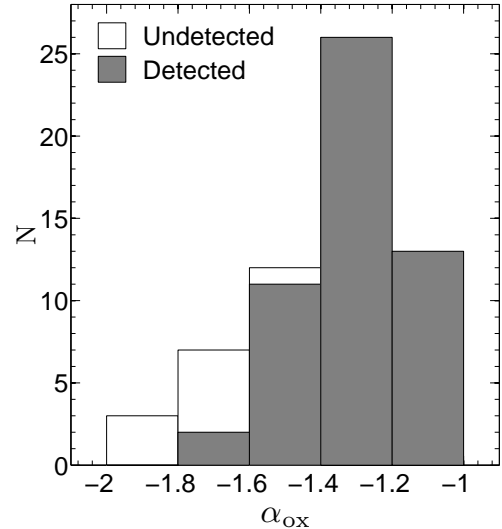}
\end{center}
\figcaption{Distribution of $\alphaox$ for our sample; each bin is 
0.2 in width in $\alphaox$.  
\label{fig:histoalphaox}}
\end{figure}

As in Paper~II, we find that the ratio of X-rays to UV is higher in
our sample of IMBHs than in AGNs with higher black hole mass. The
values in our sample range from $\alphaox \approx -1.7$ to $-1$
(Figure~\ref{fig:histoalphaox}; excluding upper limits), with an
average of $-$1.33 and a standard deviation of 0.16, higher than
$\langle\alphaox\rangle= -1.47\pm0.16$ for the W12 sample, and
$\langle\alphaox\rangle= -1.56\pm0.22$ for the 87 \citet{bor92} Palomar-Green 
(PG; Schmidt \& Green 1983) QSOs
with $z < 0.5$ and $10^7\,M_\odot\lesssim\mbh\lesssim10^9\,M_\odot$
studied by \citet[we converted their definition of $\alphaox$ to the
conventions adopted in this paper]{bra00}.  On the other hand,
our values are closer to those measured for the J12 sample,
$\langle\alphaox\rangle= -1.36\pm0.14$. In general, our $\alphaox$
values are flatter than those of SDSS-selected NLS1s (Figure 3 in
Paper~II), although both samples are drawn from the same parent
sample, and are similar to those of the X-ray-selected NLS1s studied
by \citet{gru04,gru10}.  These flatter values of $\alphaox$ are
expected from the well-known correlation between $\alphaox$ and
monochromatic UV luminosity at 2500 \AA\ (\citealt{avn82, bec03,
  str05,ste06}). We note, however, that the \alphaox\ values in our
sample span a wide range. On 
average they statistically tend to lie {\it below}\ the  best-fitting 
$\alphaox - l_{2500}$ relation of \citet{ste06}.  We will return to 
the ``X-ray--weak'' sources in Section~\ref{sec:xrayweak}.

\begin{figure}
\begin{center}
\epsscale{0.95} \plotone{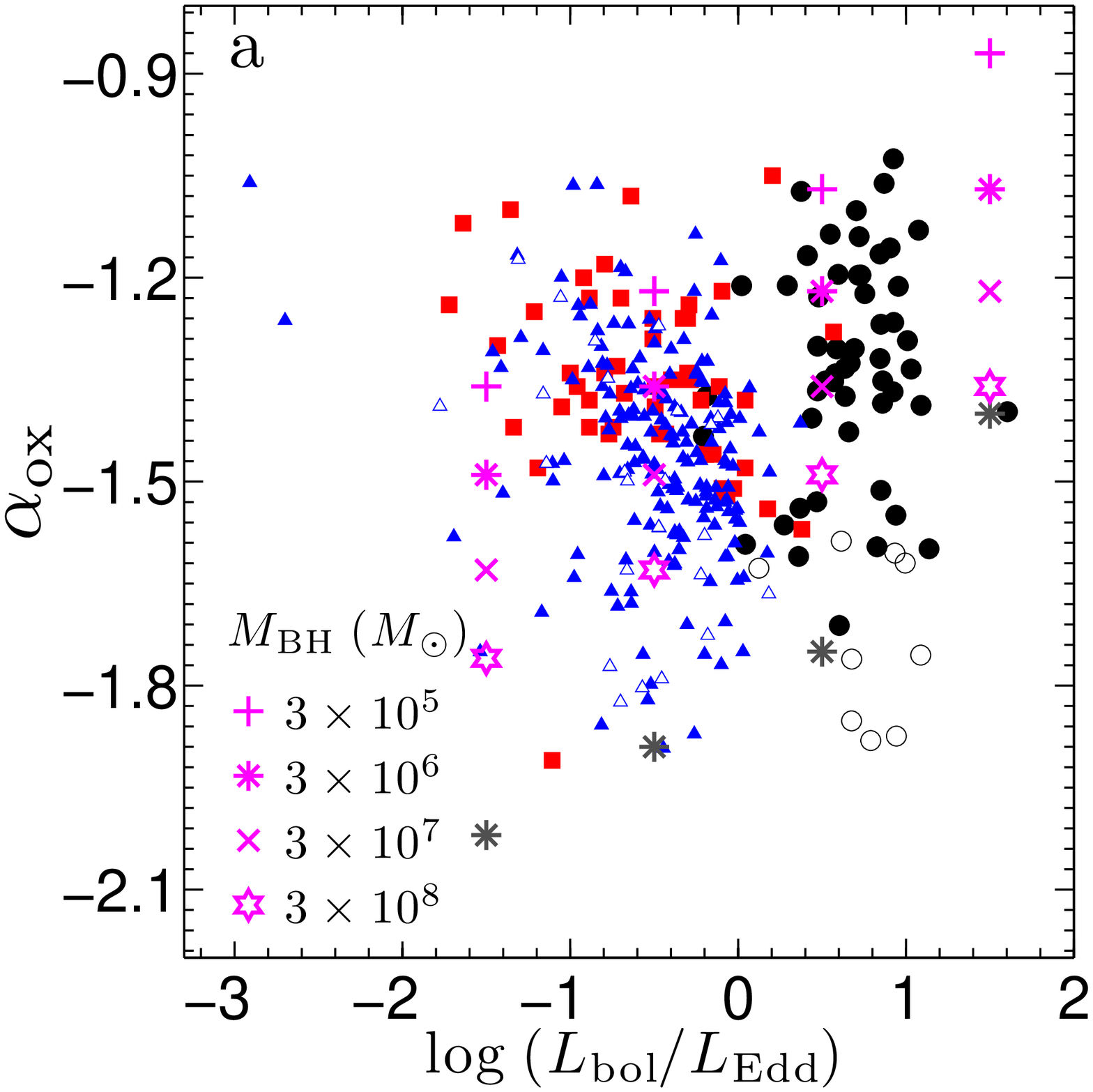} \plotone{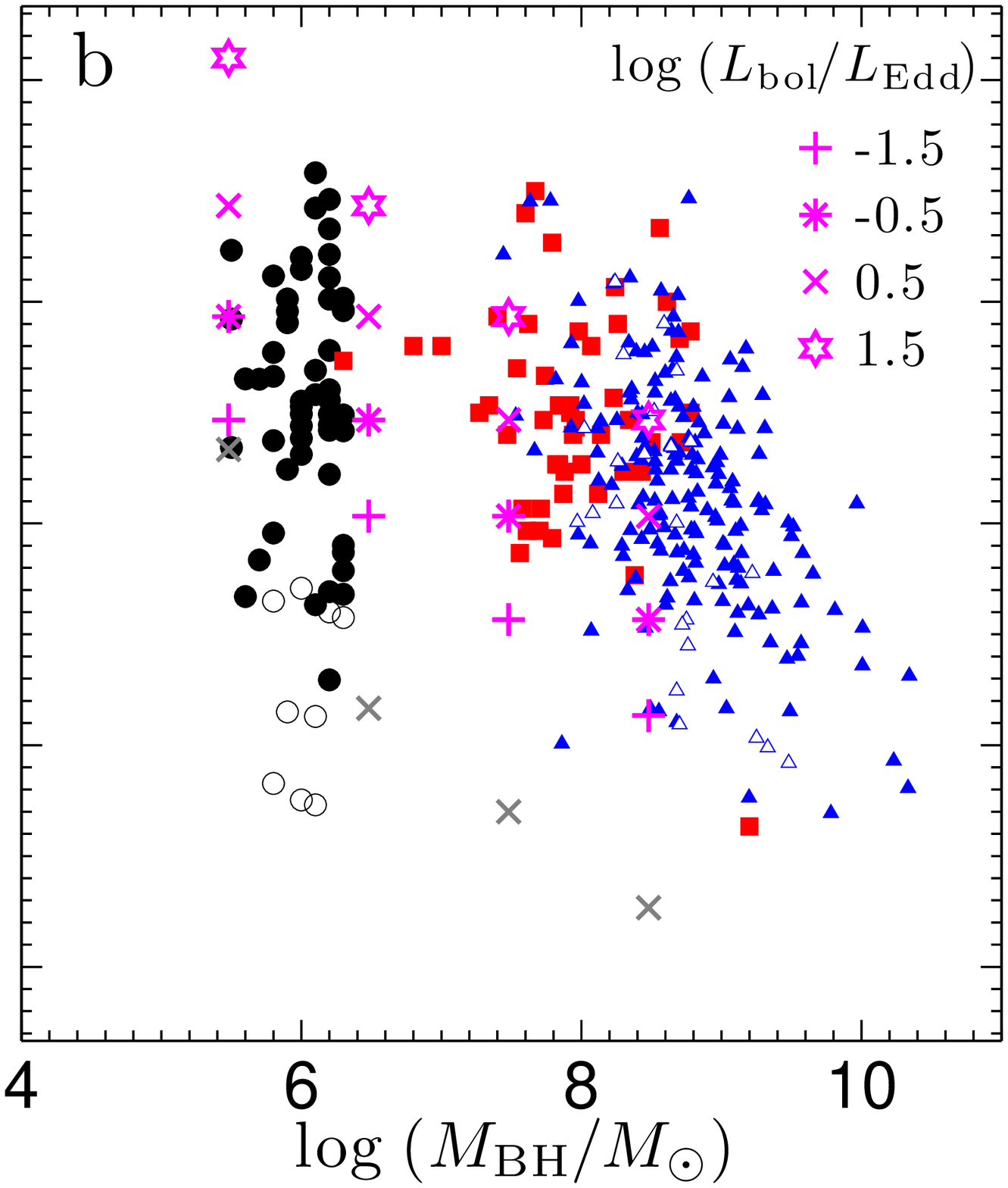}
\end{center}
\figcaption{
The dependence of $\alphaox$ on $\eddratio$ (a) and $\mbh$
(b). The black circles are from our IMBH sample, the red squares are from the \citet{jin12} sample, and the blue triangles are from \citet{wu12a}. The open symbols in each sample, whenever relevant, indicate the non-detections. \citet{don12} models are indicated by $+$, $\star$, $\times$, and hexagrams. The magenta symbols assume $R_{\rm cor}=30r_g$, while the cyan symbols assume $R_{\rm cor}=8r_g$.
\label{fig:alphaox}}
\end{figure}

We turn to putative correlations between \alphaox\ on the one hand and
\mbh\ and \eddratio\ on the other (Figure~\ref{fig:alphaox}).  $\lbol$
is derived from the specific luminosity at 5100 \AA\
($L_{5100}=l_{5100}\nu_{5100}$).  For consistency with the
J12 sample, we use an $\mbh$-dependent bolometric correction factor.
As the disk temperature increases for lower-mass BHs, less and less of
the bolometric luminosity emerges in the optical and near-UV spectrum,
leading to a mass-dependent bolometric correction. We fit a trend to
the J12 data (their table D1):
\begin{equation}
\log{\frac{L_{\rm bol}}{L_{5100}}}=-0.54\log{\mbh}+5.43.
\label{eq:bc}
\end{equation}
The bolometric correction factor used for the W12 sample (see
also \citealt{ric06}) is consistent with the J12 trend in Eq.~\ref{eq:bc} 
given the BH mass range of the W12 sample. The Eddington ratios that we derive here are
approximately $\sim 1$ dex higher than what we derive from canonical
bolometric corrections (GH07), and these rather extreme values lead to
super-Eddington accretion in our sources. However, since $\eddratio$
for both comparison samples is derived based on SED model fitting with
similar bolometric correction factors, our approach is necessary in
order to make the comparison consistent. In addition, as discussed in
Section~\ref{sec:introduction}, we have some confidence in our black
hole mass estimates on average based on the properties of their host galaxies
despite the fact that they are derived through indirect methods. We
also note that our conclusions would not change if we adopted our
standard bolometric corrections, since we find no correlation between
\alphaox\ and \eddratio\ in either case.

Consistent with Papers~I and II, $\alphaox$
shows no clear dependence on $\eddratio$. The lack of correlation has
been seen by many authors recently, spanning a wide range in redshift,
\mbh, and luminosity \citep[][]{she08,vas09,jin12,wu12a}. On the other
hand, there is a hint of a trend between \mbh\ and \alphaox.  In
particular, there is a dearth of objects with high values of \mbh\ and
\alphaox. Figure~\ref{fig:alphaox} nicely emphasizes the importance of our
sample in extending the dynamic range of existing samples,
particularly in \mbh.  It is the extra decade in \mbh\ that highlights 
a weak trend between \mbh\ and \alphaox.

To guide our interpretation, we also include theoretical predictions
from the \citet{don12} model. We show tracks in \mbh\ and \eddratio,
while leaving the other parameters fixed to default parameters
following \citet{jin12}: $kT_e=0.2$~keV (electron temperature for the
soft Comptonization component), $\tau=10$ (optical depth of the soft
Comptonization component), $\Gamma=2$ (spectral index of the hard
Comptonization component), and $f_{\rm pl}=0.1$ (fraction of the power
below $R_{\rm cor}$ that is emitted in the hard Comptonization
component).  After experimenting with different choices of these
additional parameters, we find that although the exact value of
$\alphaox$ at a given $\mbh$ and $\eddratio$ is affected by their
specific values, the predicted trends between $\alphaox$ and $\mbh$ or
$\eddratio$ are robust. There is also a long tail toward X-ray
weakness in our sample.  In Section~\ref{sec:xrayweak} we will discuss
whether these are absorbed or intrinsically weak sources.  For now, we
simply highlight the grey symbols in Figure \ref{fig:alphaox}, which
have $R_{\rm cor}=8R_g$ rather than $R_{\rm cor}=30R_g$, but otherwise are
constructed with the the same parameters.  Going from 
$R_{\rm cor}=30R_g$ to $R_{\rm cor}=8R_g$ the coronal component of the disk
shrinks, so that the Comptonization and the X-ray component of the
disk spectrum become weaker, and the disk emission is more dominated
by the (color-temperature corrected) blackbody component.

\begin{figure}
\begin{center}
\epsscale{0.95} \plotone{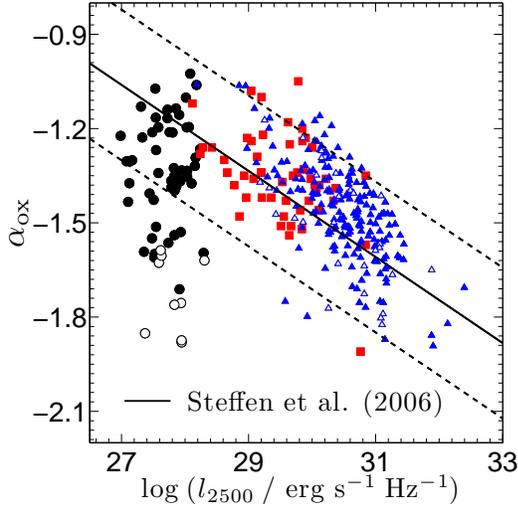}
\end{center}
\figcaption{
The dependence of $\alphaox$ on the monochromatic
luminosity at 2500 \AA ($l_{2500}$). The black circles are from our IMBH sample, the red squares are from the \citet{jin12} sample, and the blue triangles are from \citet{wu12a}. The open symbols in each sample, whenever relevant, indicate the non-detections. The 
solid line is the best-fit relation from \citet{ste06}, and the dashed lines indicate the $1\sigma$ deviations.
\label{fig:alphaoxl2500}}
\end{figure}

In the \citet{don12} models, changes in \alphaox\ are driven predominantly by
changes in the accretion disk temperature, which preferentially move
the UV luminosity up and down as the peak BBB temperature approaches
and recedes from $2500$ \AA \footnote{We note that from studying 11 ultraluminous X-ray sources,
\citet{kaj09} reached the conclusion that some super-Eddington systems
appear to have lower thermal-component temperatures as they get
brighter, while sub-Eddington systems have their thermal components
get hotter as they get brighter. Thus the assumption that the disk structure 
stays the same in sub-Eddington and super-Eddington systems may be 
oversimplified.}. Using \alphaox\, we are not very
sensitive to intrinsic changes in the X-ray SED with \mbh\ and
\eddratio\ \citep{vas09,jin12}, but we can detect changes in the BBB
temperature.  At fixed \eddratio, the BBB temperature decreases
systematically with \mbh, leading to higher (i.e. more positive) values of \alphaox\ at
lower \mbh.  The trend is highlighted clearly in the models but is
apparent in the data as well, with the mean $\alphaox$ dropping from
$-$1.33 at $\mbh \approx 10^6-10^7\,M_\odot$, to $-1.35$ at
$\mbh \approx 10^7-10^8\,M_\odot$, $-1.42$ at 
$\mbh \approx 10^8-10^8\,M_\odot$, and $-1.56$ at
$\mbh \approx 10^9-10^{10}\,M_\odot$.  The scatter is obviously
very large, presumably due to a spread in luminosity, absorption, and
corona properties at fixed \mbh.  In Figure~\ref{fig:alphaoxl2500}, we show the correlation between \alphaox\ and $l_{2500}$.  The BH mass is calculated using the same luminosity that is the independent variable in Figure~\ref{fig:alphaoxl2500}.  In both cases the correlation is weak --- the linear correlation coefficient for
\alphaox$-\log{\mbh}$ is 0.41, while for \alphaox$-\log{l_{2500}}$ it is 0.45.
However, we suggest that the correlation between \mbh\ and \alphaox\
may be the more fundamental one, arising simply from changes in the
accretion disk temperature.  We suggest that the
\alphaox$-l_{2500}$ correlation appears so strong because there is a
narrow range in \eddratio\ at a given \mbh. 

\section{X-ray--weak Sources}\label{sec:xrayweak}

\begin{figure}
\begin{center}
\epsscale{0.95} \plotone{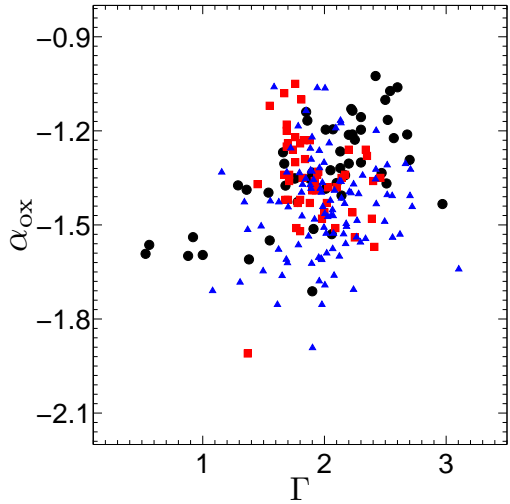}
\end{center}
\figcaption{The dependence of $\alphaox$ on the X-ray photon index. The black circles are from our IMBH sample, the red squares are from the \citet{jin12} sample, and the blue triangles are from \citet{wu12a}.
\label{fig:alphaoxgamma}}
\end{figure}

As we have seen in the above section, our sample spans a wide range in
$\alphaox$ but there is a clear excess of points that fall systematically 
below the low-luminosity extrapolation of the $\alphaox - l_{2500}$ 
relation of \citet{ste06} defined by more luminous sources.  In other words, 
some IMBHs appear to be suppressed in X-rays (relative to the UV).
Heavy obscuration provides a
simple explanation for the X-ray--weak sources. Indeed, in Figure
\ref{fig:alphaoxgamma} we see a trend toward flat $\Gamma$ values in the 
X-ray--weak sources, implicating absorption.  

\begin{figure}
\begin{center}
\epsscale{0.95} \plotone{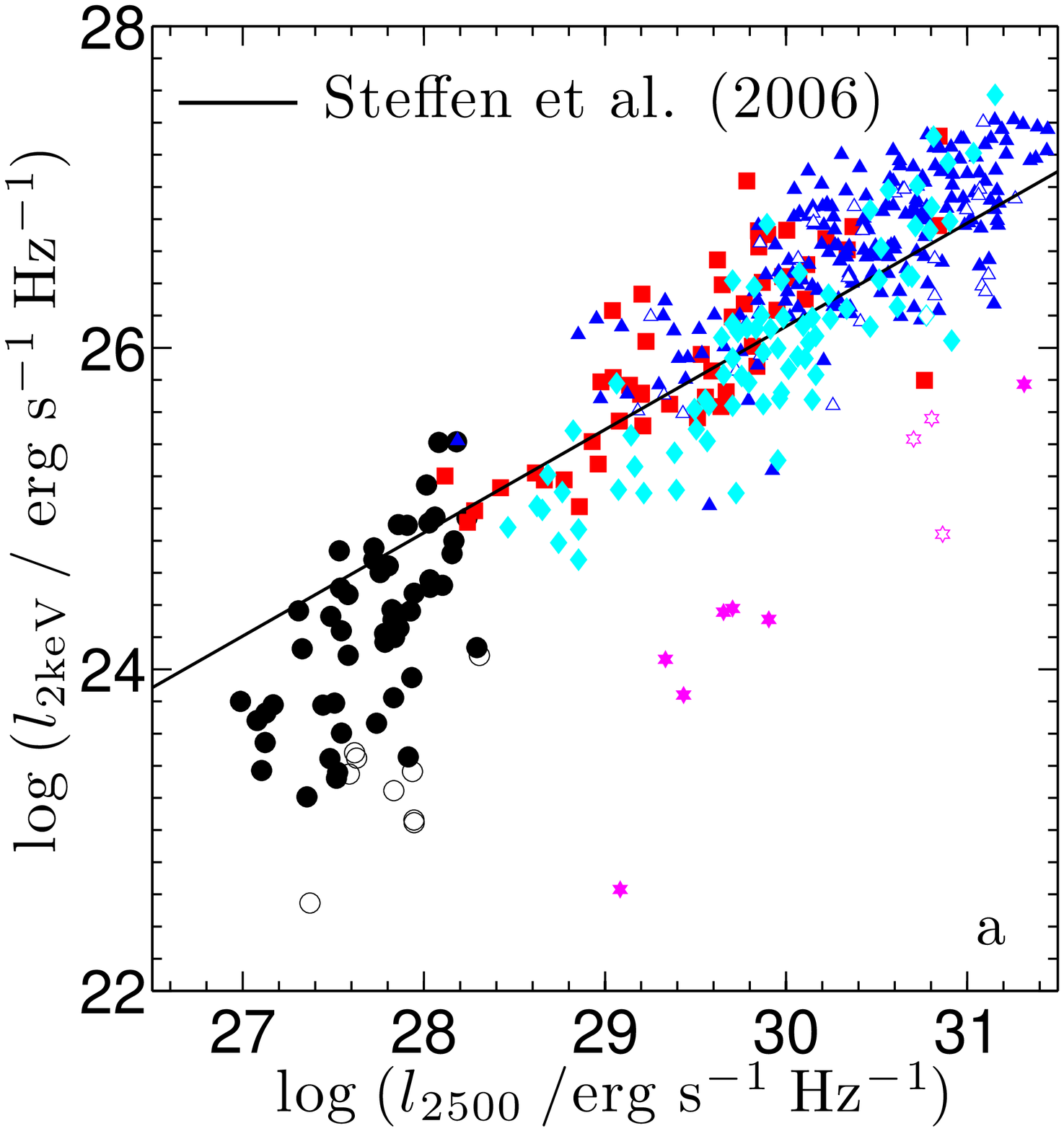} \plotone{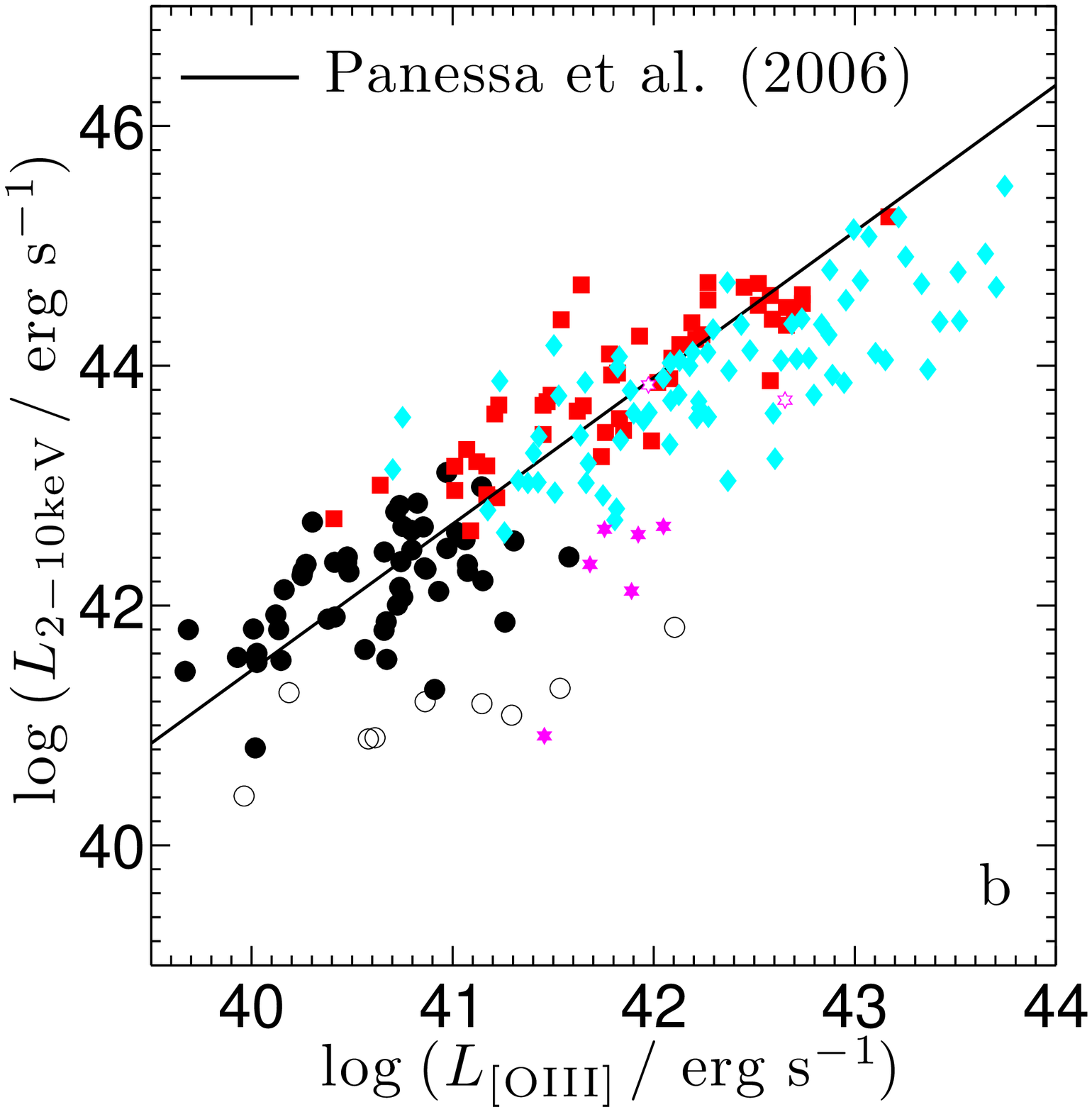}
\end{center}
\figcaption{$l_{2 \rm~keV}$ as a function of $l_{2500}$ (a), and $L_{2-10 \, \rm keV}$ as a function of $L_{\rm [O~III]}$ (b). The 
solid lines are the best-fit relations from past literature. The black circles are from our IMBH sample, the red squares are from the \citet{jin12} sample, and the blue triangles are from \citet{wu12a}. For the IMBHs, $L_{2-10 \, \rm keV}$ is 
calculated from $L_{2-8 \, \rm keV}$ and $\gammahr$. We also include the PG QSOs from 
Brandt et al. (2000; light blue diamonds; soft X-ray-weak objects shown in magenta hexagrams), using values of $L_{2-10 \, \rm keV}$ estimated from 
$L_{5100}$, $\alphaox$, an assumed optical-UV spectrum $f_\nu\propto 
\nu^{-0.44}$ \citep{van01}, and $\Gamma_h = 1.9$ (Piconcelli et al. 2005); 
$L_{\rm [O~III]}$ was derived from the \oiii\ equivalent widths of Boroson \& 
Green (1992) and the optical spectrophotometry of Neugebauer et al. (1987). 
Open symbols indicate the X-ray non-detections.
\label{fig:xrayweak}}
\end{figure}

However, not all the sources follow this trend.  There are a few
objects in Figure~\ref{fig:alphaoxgamma} that have typical
sample-averaged $\Gamma$ values ($\sim2$) but low $\alphaox$
values. They raise the intriguing possibility that some of these sources
are intrinsically X-ray--weak. In Figure~\ref{fig:xrayweak}a, we show
the relation between 2~keV and UV luminosity in a slightly different
way.  For reference we have added the sample of $z < 0.5$ PG QSOs from 
Brandt et al. (2000; light blue diamonds), highlighting in magenta hexagrams 
the subset of objects that they dub X-ray--weak quasars.
These comprise broad absorption-line quasars, and deep X-ray
observations confirm that their X-ray weakness is caused by absorption
\citep[e.g.,][]{gal01,bri04,sch05,bal08,gal11}.  A minority, however,
appear to be genuinely X-ray--weak
\citep[e.g.,][]{ris01,sab01,lei07a,min09,bal11}, and this unusual
spectral departure has been suggested to be a possible signature of
accretion systems with high Eddington ratios.  PHL 1811 is the prototype: not 
only are its X-rays unabsorbed, but they vary on a short time scale, indicating 
a compact size. Such objects are quite rare among optically selected samples 
of QSOs, comprising less than 2\% of QSOs in SDSS \citep{gibson08}.

In the absence of UV spectroscopy or X-ray variability information, it
is difficult to say for certain whether any of our sources are
intrinsically X-ray--weak.  In view of the bias toward high accretion
rates in IMBHs, intrinsic weakness is an attractive possibility to account for the
large fraction of X-ray--weak sources in our sample. As mentioned
before, type~1 IMBHs, as a class, show little direct evidence for
additional absorption above the Galactic value.  A few of the
X-ray--weak sources do have exceptionally hard X-ray spectra, with
$\Gamma \approx 1$, but the majority, in fact, have photon indices
that span the same range as observed for the X-ray ``normal'' sources
($\Gamma_s \approx 1.5-3$; Figure~\ref{fig:alphaoxl2500}).  Also,
notably, Brandt et al.'s X-ray--weak PG QSOs form a disjoint group
significantly offset from the main population in
Figure~\ref{fig:xrayweak}a (average UV/X-ray ratio lower by a factor
$\sim 60$), whereas, by contrast, our X-ray--weak sources seem to join
more or less smoothly in a continuous sequence with the ``normal''
sources (average UV/X-ray ratio lower by a factor $\sim 5$).

A similar perspective comes from comparing the X-ray luminosity with
some measure of optical line emission, as the two are strongly
correlated in unobscured AGNs
\citep[e.g.,][]{elv84,war88,mul94,ter00,ho01,hec05}.  Heavily obscured
AGNs, such as Compton-thick systems, fall off the correlation in being
underluminous in X-rays for a given optical line luminosity
\citep[e.g.,][]{bas99}. The correlation between 2--10 keV luminosity
with \oiii\ $\lambda$5007 luminosity is shown in
Figure~\ref{fig:xrayweak}b, where, for comparison, we include the
  best-fitting relation from a sample of low-luminosity Seyferts in
  \citet{pan06}.  The X-ray--weak IMBHs do tend to lie
systematically below the relation defined by the higher luminosity
sources, suggesting that absorption is likely an explanation for many 
of the sources, but perhaps not all. 

In principle, infrared data could also serve as a proxy for the bolometric luminosity of the system, to indicate whether the UV-weak sources are absorbed or intrinsically weak.  We examine the IR luminosities of our sample in the Wide-field Infrared Survey Explorer (WISE) all-sky catalog\footnote{\tt http://irsa.ipac.caltech.edu/cgi-bin/Gator/nph-dd}. All of our objects except three are detected in WISE. However, the distribution of their W1-W2 colors (mean=0.63, $\sigma=0.27$) suggests that the photometry is dominated by the light from the host galaxies in most cases \citep{ass12}. This is consistent with the beam size of the WISE observations ($6\arcsec-12\arcsec$ at various bands). Without detailed SED decomposition, it is not possible to determine the AGN luminosities from the WISE data alone.  We need to await higher resolution data.

At the moment, the true nature of the X-ray--weak sources is unclear.
We suggest two future observations that could test whether the observed 
X-ray weakness is truly intrinsic:

\begin{enumerate}
\item X-ray observations with enough counts and energy resolution to allow 
detailed spectral fitting to determine if there is evidence for intrinsic 
absorption.

\item High-resolution IR observations to check if the X-ray photons have been absorbed and 
reprocessed as IR emission by dust.
\end{enumerate}

\section{Summary}
\label{sec:summary}

We carry out a study of the X-ray properties of 49 IMBHs in active
galaxies selected from the GH07 sample using \chandra\ data. We detect 42 out of 49 objects at the SDSS position. The
main results are consistent with Greene \& Ho (2007a) and Desroches et
al. (2009).  We perform spectral analysis of 10 objects with sufficient
counts, and all of them are consistent with a Galactic absorbed power-law
model. Spectral fitting with intrinsic absorption and an additional
disk blackbody component generally produce a poor result, yielding
essentially zero intrinsic column density as well as unphysical
$T_{\rm in}$. We conclude that there is no significant sign of
intrinsic absorption or a disk blackbody component in the bright
sources. For two objects with enough counts in the hard X-ray band, we
fit the spectra using an absorbed power-law model using data $\geq1.5$ keV,
then extrapolate it to the whole energy range. SDSS~J1559+3501 shows 
good agreement between the extrapolation and the spectra at lower
energy, while SDSS~J0903+0457 shows a prominent soft excess. We
conclude that with enough counts, both situations are possible in the
low-mass regime. 

The $\Gamma$ values that we measure for the IMBH sample occupy the
same range (0.5--3) as their more luminous counterparts.  However,
given the limited spectral range of our 2~ks observations, we caution
against detailed comparisons between our inferred slopes and those
based on detailed spectroscopy with energy sensitivity at $\gtrsim 10$
keV. Exploring the true range in $\Gamma$ requires deeper spectroscopy.

Our sample contains significantly flatter values of $\alphaox$ than in more 
luminous PG QSOs \citep{bra00} and SDSS-selected NLS1s (Paper~II).  Our values 
of $\alphaox$ generally agree with the low-luminosity extrapolation of the 
well-known $\alphaox-l_{2500}$ correlation, but, at a given UV 
luminosity, there is a tendency for IMBHs to be more X-ray--weak.  While 
absorption may be partly responsible for this trend, we suggest the intriguing 
possibility that some of the objects may be intrinsically X-ray--weak, as 
previously suggested for AGNs with high \eddratio.  Deeper X-ray follow-up of 
the X-ray--weak sources would determine whether some significant fraction of 
our sample is actually in a mode of accretion with little to no X-ray corona.  

While \alphaox\ shows no correlation with \eddratio, we
do find a mild correlation between \alphaox\ and \mbh, in just the
sense expected by disk models.  As \mbh\ drops, the disk temperature
increases, and the sources grow fainter in the near-UV, thus
increasing the level of X-rays relative to the UV flux.  Indeed, this
correlation between \alphaox\ and \mbh\ may drive the well-known
correlation between $l_{2500}$ and \alphaox. Considering that
extinction, variations in the corona, and uncertainties in the \mbh\
measurements all wash out this trend, the large dynamic range in \mbh\
afforded by our sample is essential to detect any correlation at all.

\section*{Acknowledgments}
We thank Xiao-Bo Dong and Jesper Rasmussen for useful discussions, 
Louis-Benoit Desroches for providing part of the comparison data, and Craig 
Gordon from the XSPEC online help team for technical support. The authors are 
also grateful to the anonymous referee who helped improve the manuscript. 
This work was funded through SAO grant 11700259.

\clearpage
\LongTables
\begin{deluxetable}{cccccccccccc}
\tablewidth{0pt}
\tabletypesize{\scriptsize}
\tablecaption{X-ray Properties}
\tablehead{\colhead{SDSS ID} & \colhead{$D_L$} & \colhead{$\log N_{\rm H}$} & \colhead{S/N} & \colhead{$C_s$} & \colhead{$C_h$} & \colhead{$\Gamma_{\rm HR}$} & \colhead{$\log f_s$} & \colhead{$\log f_h$} & \colhead{$\log L_s$} & \colhead{$\log L_h$} & \colhead{$\alpha_{\rm ox}$} \\ 
\colhead{} & \colhead{(Mpc)} & \colhead{(cm$^{-2}$)} & \colhead{} & \colhead{(count s$^{-1}$)} & \colhead{(count s$^{-1}$)} & \colhead{} & \colhead{(erg s$^{-1}$ cm$^{-2}$)} & \colhead{(erg s$^{-1}$ cm$^{-2}$)} & \colhead{(erg s$^{-1}$)} & \colhead{(erg s$^{-1}$)} & \colhead{} \\
\colhead{(1)} & \colhead{(2)} & \colhead{(3)} & \colhead{(4)} & \colhead{(5)} & \colhead{(6)} & \colhead{(7)} & \colhead{(8)} & \colhead{(9)} & \colhead{(10)} & \colhead{(11)} & \colhead{(12)}}
\startdata
J0001$-$1002 & 215 & 20.48 &  3.7 & $0.0061\pm{0.0018}$ & $0.0045\pm{0.0015}$ & $1.29\pm{0.33}$ & $-13.60^{+0.12}_{-0.15}$ & $-13.05^{+0.16}_{-0.22}$ & $41.15^{+0.12}_{-0.15}$ & $41.70^{+0.16}_{-0.22}$ &$-$1.37\\
J0228$-$0902 & 322 & 20.55 &  7.4 & $0.0280\pm{0.0039}$ & $0.0060\pm{0.0018}$ & $2.22\pm{0.27}$ & $-12.91^{+0.07}_{-0.07}$ & $-13.02^{+0.14}_{-0.18}$ & $42.19^{+0.07}_{-0.07}$ & $42.08^{+0.14}_{-0.18}$ &$-$1.13\\
J0304$+$0028 & 194 & 20.86 & 25.5 & $0.2599\pm{0.0115}$ & $0.0788\pm{0.0063}$ & $2.07\pm{0.07}$ & $-11.90^{+0.02}_{-0.02}$ & $-11.89^{+0.04}_{-0.04}$ & $42.76^{+0.02}_{-0.02}$ & $42.77^{+0.04}_{-0.04}$ &$-$1.20\\
J0731$+$3926 & 212 & 20.79 & 18.7 & $0.1813\pm{0.0104}$ & $0.0285\pm{0.0041}$ & $2.54\pm{0.12}$ & $-12.05^{+0.03}_{-0.03}$ & $-12.37^{+0.07}_{-0.08}$ & $42.68^{+0.03}_{-0.03}$ & $42.36^{+0.07}_{-0.08}$ &$-$1.07\\
J0806$+$2420 & 182 & 20.61 &  6.6 & $0.0191\pm{0.0031}$ & $0.0076\pm{0.0020}$ & $1.78\pm{0.22}$ & $-13.08^{+0.07}_{-0.08}$ & $-12.88^{+0.12}_{-0.15}$ & $41.52^{+0.07}_{-0.08}$ & $41.72^{+0.12}_{-0.15}$ &$-$1.43\\
J0809$+$4416 & 238 & 20.70 & 11.6 & $0.0569\pm{0.0054}$ & $0.0173\pm{0.0030}$ & $2.01\pm{0.15}$ & $-12.59^{+0.04}_{-0.05}$ & $-12.54^{+0.08}_{-0.10}$ & $42.25^{+0.04}_{-0.05}$ & $42.30^{+0.08}_{-0.10}$ &$-$1.20\\
J0816$+$2506 & 324 & 20.59 &  5.5 & $0.0151\pm{0.0029}$ & $0.0070\pm{0.0020}$ & $1.66\pm{0.26}$ & $-13.19^{+0.08}_{-0.10}$ & $-12.90^{+0.14}_{-0.17}$ & $41.92^{+0.08}_{-0.10}$ & $42.21^{+0.14}_{-0.17}$ &$-$1.27\\
J0824$+$2959 & 109 & 20.58 &  6.8 & $0.0126\pm{0.0025}$ & $0.0166\pm{0.0029}$ & $0.88\pm{0.20}$ & $-13.28^{+0.08}_{-0.10}$ & $-12.43^{+0.09}_{-0.11}$ & $40.88^{+0.08}_{-0.10}$ & $41.73^{+0.09}_{-0.11}$ &$-$1.60\\
J0852$+$5228 & 286 & 20.40 &  7.7 & $0.0250\pm{0.0036}$ & $0.0097\pm{0.0022}$ & $1.76\pm{0.20}$ & $-12.98^{+0.06}_{-0.07}$ & $-12.77^{+0.11}_{-0.14}$ & $42.01^{+0.06}_{-0.07}$ & $42.23^{+0.11}_{-0.14}$ &$-$1.35\\
J0903$+$0457 & 250 & 20.62 & 36.9 & $0.5916\pm{0.0179}$ & $0.1003\pm{0.0074}$ & $2.42\pm{0.06}$ & $-11.57^{+0.02}_{-0.02}$ & $-11.82^{+0.04}_{-0.04}$ & $43.31^{+0.02}_{-0.02}$ & $43.06^{+0.04}_{-0.04}$ &$-$1.03\\
J0924$+$5607 & 107 & 20.42 &  6.4 & $0.0160\pm{0.0029}$ & $0.0103\pm{0.0023}$ & $1.38\pm{0.21}$ & $-13.18^{+0.08}_{-0.09}$ & $-12.70^{+0.11}_{-0.13}$ & $40.96^{+0.08}_{-0.09}$ & $41.44^{+0.11}_{-0.13}$ &$-$1.61\\
J0933$+$5347 & 251 & 20.23 & 11.6 & $0.0600\pm{0.0055}$ & $0.0080\pm{0.0020}$ & $2.51\pm{0.19}$ & $-12.59^{+0.04}_{-0.05}$ & $-12.92^{+0.11}_{-0.14}$ & $42.29^{+0.04}_{-0.05}$ & $41.96^{+0.11}_{-0.14}$ &$-$1.37\\
J0941$+$0324 & 268 & 20.58 & 19.0 & $0.1517\pm{0.0088}$ & $0.0296\pm{0.0039}$ & $2.30\pm{0.11}$ & $-12.17^{+0.03}_{-0.03}$ & $-12.33^{+0.06}_{-0.07}$ & $42.77^{+0.03}_{-0.03}$ & $42.60^{+0.06}_{-0.07}$ &$-$1.20\\
J0953$+$5627 & 294 & 20.04 &  --- & $<$0.0031 & $<$0.0008 & --- & $<-$13.90 & $<-$13.90 & $<$41.12 & $<$41.12 & $<-$1.63\\
J0954$+$4032 & 299 & 20.19 &  9.4 & $0.0369\pm{0.0045}$ & $0.0152\pm{0.0029}$ & $1.68\pm{0.17}$ & $-12.82^{+0.05}_{-0.06}$ & $-12.56^{+0.09}_{-0.11}$ & $42.21^{+0.05}_{-0.06}$ & $42.47^{+0.09}_{-0.11}$ &$-$1.37\\
J1035$+$0734 & 299 & 20.46 &  7.3 & $0.0244\pm{0.0035}$ & $0.0096\pm{0.0022}$ & $1.75\pm{0.19}$ & $-12.99^{+0.06}_{-0.07}$ & $-12.77^{+0.11}_{-0.13}$ & $42.04^{+0.06}_{-0.07}$ & $42.26^{+0.11}_{-0.13}$ &$-$1.35\\
J1058$+$4825 & 327 & 20.09 &  --- & $<$0.0031 & $<$0.0008 & --- & $<-$13.90 & $<-$13.90 & $<$41.21 & $<$41.21 & $<-$1.61\\
J1105$+$5941 & 146 & 19.76 & 13.6 & $0.0729\pm{0.0063}$ & $0.0351\pm{0.0044}$ & $1.54\pm{0.11}$ & $-12.54^{+0.04}_{-0.04}$ & $-12.19^{+0.05}_{-0.06}$ & $41.87^{+0.04}_{-0.04}$ & $42.23^{+0.05}_{-0.06}$ &$-$1.40\\
J1125$+$0220 & 213 & 20.58 &  9.5 & $0.0450\pm{0.0050}$ & $0.0115\pm{0.0025}$ & $2.10\pm{0.18}$ & $-12.70^{+0.05}_{-0.06}$ & $-12.73^{+0.10}_{-0.12}$ & $42.04^{+0.05}_{-0.06}$ & $42.01^{+0.10}_{-0.12}$ &$-$1.37\\
J1136$+$4246 & 317 & 20.34 & 12.2 & $0.0601\pm{0.0055}$ & $0.0202\pm{0.0032}$ & $1.85\pm{0.14}$ & $-12.60^{+0.04}_{-0.04}$ & $-12.46^{+0.08}_{-0.09}$ & $42.48^{+0.04}_{-0.04}$ & $42.62^{+0.08}_{-0.09}$ &$-$1.14\\
J1137$+$4113 & 319 & 20.32 & 10.5 & $0.0518\pm{0.0052}$ & $0.0104\pm{0.0023}$ & $2.23\pm{0.17}$ & $-12.66^{+0.05}_{-0.05}$ & $-12.79^{+0.10}_{-0.13}$ & $42.43^{+0.05}_{-0.05}$ & $42.30^{+0.10}_{-0.13}$ &$-$1.21\\
J1143$+$5500 & 117 & 20.03 &  --- & $<$0.0033 & $<$0.0008 & --- & $<-$13.88 & $<-$13.88 & $<$40.35 & $<$40.35 & $<-$1.85\\
J1153$+$4612 & 104 & 20.30 & 26.7 & $0.2829\pm{0.0119}$ & $0.0715\pm{0.0060}$ & $2.05\pm{0.07}$ & $-11.93^{+0.02}_{-0.02}$ & $-11.93^{+0.04}_{-0.05}$ & $42.19^{+0.02}_{-0.02}$ & $42.19^{+0.04}_{-0.05}$ &$-$1.33\\
J1215$+$0148 & 317 & 20.27 &  6.5 & $0.0210\pm{0.0033}$ & $0.0056\pm{0.0017}$ & $2.01\pm{0.25}$ & $-13.06^{+0.07}_{-0.08}$ & $-13.03^{+0.14}_{-0.18}$ & $42.02^{+0.07}_{-0.08}$ & $42.05^{+0.14}_{-0.18}$ &$-$1.38\\
J1223$+$5814 &  61 & 20.09 & 19.5 & $0.1541\pm{0.0092}$ & $0.0642\pm{0.0059}$ & $1.67\pm{0.08}$ & $-12.21^{+0.03}_{-0.03}$ & $-11.94^{+0.05}_{-0.05}$ & $41.45^{+0.03}_{-0.03}$ & $41.72^{+0.05}_{-0.05}$ &$-$1.30\\
J1227$+$6306 & 350 & 20.27 &  4.4 & $0.0126\pm{0.0026}$ & $0.0027\pm{0.0012}$ & $2.17\pm{0.38}$ & $-13.28^{+0.09}_{-0.11}$ & $-13.36^{+0.20}_{-0.30}$ & $41.89^{+0.09}_{-0.11}$ & $41.81^{+0.20}_{-0.30}$ &$-$1.34\\
J1254$+$4628 & 272 & 20.15 &  2.9 & $0.0071\pm{0.0020}$ & $0.0005\pm{0.0005}$ & $2.97\pm{0.86}$ & $\sim -13.51$ & $\sim -14.17$ & $\sim 41.44$ & $\sim 40.78$ & $-$1.43\\
J1308$+$5052 & 239 & 20.07 & 15.2 & $0.0950\pm{0.0069}$ & $0.0303\pm{0.0039}$ & $1.86\pm{0.11}$ & $-12.41^{+0.03}_{-0.03}$ & $-12.28^{+0.06}_{-0.07}$ & $42.42^{+0.03}_{-0.03}$ & $42.55^{+0.06}_{-0.07}$ &$-$1.17\\
J1313$+$0519 & 214 & 20.34 & 15.9 & $0.1159\pm{0.0076}$ & $0.0156\pm{0.0028}$ & $2.52\pm{0.14}$ & $-12.30^{+0.03}_{-0.03}$ & $-12.63^{+0.08}_{-0.10}$ & $42.44^{+0.03}_{-0.03}$ & $42.11^{+0.08}_{-0.10}$ &$-$1.17\\
J1317$+$0556 & 243 & 20.36 &  4.8 & $0.0076\pm{0.0020}$ & $0.0091\pm{0.0022}$ & $0.92\pm{0.26}$ & $-13.51^{+0.10}_{-0.13}$ & $-12.70^{+0.12}_{-0.15}$ & $41.34^{+0.10}_{-0.13}$ & $42.15^{+0.12}_{-0.15}$ &$-$1.54\\
J1319$+$1056 & 286 & 20.27 &  --- & $<$0.0033 & $<$0.0008 & --- & $<-$13.86 & $<-$13.86 & $<$41.13 & $<$41.13 & $<-$1.76\\
J1322$+$4226 & 332 & 20.09 &  5.3 & $0.0112\pm{0.0024}$ & $0.0071\pm{0.0019}$ & $1.36\pm{0.25}$ & $-13.35^{+0.09}_{-0.11}$ & $-12.86^{+0.13}_{-0.16}$ & $41.77^{+0.09}_{-0.11}$ & $42.27^{+0.13}_{-0.16}$ &$-$1.39\\
J1347$+$4743 & 285 & 20.23 & 11.5 & $0.0697\pm{0.0062}$ & $0.0099\pm{0.0023}$ & $2.47\pm{0.19}$ & $-12.53^{+0.04}_{-0.05}$ & $-12.83^{+0.11}_{-0.13}$ & $42.46^{+0.04}_{-0.05}$ & $42.16^{+0.11}_{-0.13}$ &$-$1.33\\
J1433$+$5258 & 208 & 20.15 &  2.6 & $0.0022\pm{0.0011}$ & $0.0043\pm{0.0015}$ & $0.53\pm{0.45}$ & $-14.07^{+0.18}_{-0.30}$ & $-12.97^{+0.19}_{-0.25}$ & $40.65^{+0.18}_{-0.30}$ & $41.74^{+0.19}_{-0.25}$ &$-$1.59\\
J1441$-$0235 & 194 & 20.64 &  --- & $<$0.0034 & $<$0.0009 & --- & $<-$13.83 & $<-$13.83 & $<$40.83 & $<$40.83 & $<-$1.88\\
J1534$+$0408 & 172 & 20.63 &  8.3 & $0.0352\pm{0.0044}$ & $0.0049\pm{0.0016}$ & $2.57\pm{0.27}$ & $-12.79^{+0.06}_{-0.07}$ & $-13.14^{+0.15}_{-0.20}$ & $41.77^{+0.06}_{-0.07}$ & $41.41^{+0.15}_{-0.20}$ &$-$1.22\\
J1537$+$3122 & 248 & 20.37 &  --- & $<$0.0034 & $<$0.0008 & --- & $<-$13.85 & $<-$13.85 & $<$41.02 & $<$41.02 & $<-$1.76\\
J1542$+$3100 & 304 & 20.38 & 16.1 & $0.1130\pm{0.0077}$ & $0.0228\pm{0.0034}$ & $2.23\pm{0.12}$ & $-12.32^{+0.03}_{-0.03}$ & $-12.44^{+0.07}_{-0.08}$ & $42.73^{+0.03}_{-0.03}$ & $42.61^{+0.07}_{-0.08}$ &$-$1.14\\
J1543$+$0307 & 291 & 20.75 &  9.8 & $0.0379\pm{0.0044}$ & $0.0140\pm{0.0027}$ & $1.88\pm{0.17}$ & $-12.76^{+0.05}_{-0.06}$ & $-12.62^{+0.09}_{-0.11}$ & $42.25^{+0.05}_{-0.06}$ & $42.39^{+0.09}_{-0.11}$ &$-$1.35\\
J1559$+$3501 & 134 & 20.32 & 40.4 & $0.6521\pm{0.0183}$ & $0.1502\pm{0.0088}$ & $2.13\pm{0.05}$ & $-11.56^{+0.01}_{-0.01}$ & $-11.61^{+0.03}_{-0.03}$ & $42.78^{+0.01}_{-0.01}$ & $42.72^{+0.03}_{-0.03}$ &$-$1.27\\
J1618$-$0020 & 252 & 20.85 &  6.9 & $0.0254\pm{0.0036}$ & $0.0061\pm{0.0018}$ & $2.25\pm{0.24}$ & $-12.91^{+0.07}_{-0.08}$ & $-13.02^{+0.13}_{-0.17}$ & $41.98^{+0.07}_{-0.08}$ & $41.87^{+0.13}_{-0.17}$ &$-$1.23\\
J1624$-$0054 & 204 & 20.87 &  5.5 & $0.0164\pm{0.0029}$ & $0.0046\pm{0.0015}$ & $2.14\pm{0.28}$ & $-13.10^{+0.08}_{-0.10}$ & $-13.12^{+0.15}_{-0.20}$ & $41.60^{+0.08}_{-0.10}$ & $41.57^{+0.15}_{-0.20}$ &$-$1.41\\
J1626$+$3502 & 148 & 20.16 &  5.1 & $0.0123\pm{0.0025}$ & $0.0061\pm{0.0018}$ & $1.55\pm{0.02}$ & $-13.30^{+0.08}_{-0.10}$ & $-12.94^{+0.11}_{-0.15}$ & $41.12^{+0.08}_{-0.10}$ & $41.48^{+0.11}_{-0.15}$ &$-$1.55\\
J1632$-$0028 & 189 & 20.58 &  5.1 & $0.0123\pm{0.0025}$ & $0.0041\pm{0.0015}$ & $1.91\pm{0.30}$ & $-13.27^{+0.09}_{-0.11}$ & $-13.15^{+0.16}_{-0.22}$ & $41.37^{+0.09}_{-0.11}$ & $41.48^{+0.16}_{-0.22}$ &$-$1.51\\
J1632$+$2437 & 314 & 20.79 &  --- & $<$0.0032 & $<$0.0009 & --- & $<-$13.83 & $<-$13.83 & $<$41.24 & $<$41.24 & $<-$1.59\\
J1656$+$3714 & 277 & 20.26 &  5.1 & $0.0141\pm{0.0027}$ & $0.0035\pm{0.0013}$ & $2.06\pm{0.32}$ & $-13.23^{+0.08}_{-0.10}$ & $-13.24^{+0.17}_{-0.24}$ & $41.74^{+0.08}_{-0.10}$ & $41.73^{+0.17}_{-0.24}$ & $-$1.53\\
J2058$-$0650 & 329 & 20.70 & 10.7 & $0.0553\pm{0.0055}$ & $0.0142\pm{0.0028}$ & $2.13\pm{0.16}$ & $-12.60^{+0.05}_{-0.05}$ & $-12.64^{+0.09}_{-0.11}$ & $42.52^{+0.05}_{-0.05}$ & $42.48^{+0.09}_{-0.11}$ & $-$1.32\\
J2348$-$0912 & 348 & 20.45 &  2.6 & $0.0022\pm{0.0011}$ & $0.0043\pm{0.0015}$ & $0.56\pm{0.44}$ & $-14.06^{+0.18}_{-0.30}$ & $-12.98^{+0.19}_{-0.24}$ & $41.11^{+0.18}_{-0.30}$ & $42.19^{+0.19}_{-0.24}$ & $-$1.56\\
J2358$+$0035 & 338 & 20.51 &  7.8 & $0.0312\pm{0.0040}$ & $0.0036\pm{0.0014}$ & $2.68\pm{0.30}$ & $-12.85^{+0.06}_{-0.07}$ & $-13.29^{+0.17}_{-0.23}$ & $42.29^{+0.06}_{-0.07}$ & $41.85^{+0.17}_{-0.23}$ & $-$1.21\\

\enddata
\tablecomments{Col. (1): Abbreviated SDSS name. Col. (2): Luminosity distance, calculated using the observed SDSS redshift and our adopted cosmology. Col. (3): Column density of neutral hydrogen, from \citet{dic90}, calculated using WebPIMMS. Col. (4): Signal-to-noise ratio of the detected sources. Col. (5): Count rate in the 0.5--2~keV band. Col. (6): Count rate in the 2--8~keV band. Col. (7): Photon index, where $N(E) \propto E^{-\Gamma_{\rm HR}}$, estimated from the hardness ratio (see text). Col. (8): Flux in the 0.5--2~keV band. Col. (9): Flux in the 2--8~keV band. Col. (10): Luminosity in the 0.5--2~keV band. Col. (11): Luminosity in the 2--8~keV band.  Col. (12): Spectral index between 2500~\AA\ and 2~keV, such that $\alpha_{\rm ox} = -0.3838 \log (f_{\rm 2500~\AA}/f_{\rm 2 ~keV})$; see text for details.}
\label{tab:x-rayproperty}
\end{deluxetable}
\clearpage

\begin{deluxetable}{lccccc}
\tabletypesize{\scriptsize}
\tablewidth{0pc}
\tablecaption{Results of Spectral Fits}
\tablehead{\colhead{SDSS ID} & \colhead{Energy Range} & \colhead{$\Gamma_s$} & \colhead{Normalization} & \colhead{$\chi_\nu^2$} & \colhead{dof} \\
\colhead{} & \colhead{(keV)} & \colhead{} & \colhead{(10$^{-4}$ photon s$^{-1}$ keV$^{-1}$)} & \colhead{} & \colhead{}\\
\colhead{(1)} & \colhead{(2)} & \colhead{(3)} & \colhead{(4)} & \colhead{(5)} & \colhead{(6)}}
\startdata
J0304$+$0028 & 0.3--5 & 1.91$\pm$0.08 &  5.35$\pm$0.47 & 0.90 & 28\\
J0731$+$3926 & 0.3--3 & 2.71$\pm$0.15 &  3.87$\pm$0.41 & 0.98 & 13\\
J0903$+$0457 & 0.3--5 & 2.75$\pm$0.06 & 11.75$\pm$0.58 & 1.67 & 50\\
J0941$+$0324 & 0.3--5 & 2.52$\pm$0.12 &  2.91$\pm$0.32 & 1.58 & 20\\
J1153$+$4612 & 0.3--5 & 2.20$\pm$0.08 &  5.39$\pm$0.40 & 1.31 & 30\\
J1223$+$5814 & 0.3--5 & 1.54$\pm$0.10 &  2.63$\pm$0.32 & 1.42 & 16\\
J1308$+$5052 & 0.5--4 & 1.51$\pm$0.20 &  1.64$\pm$0.27 & 1.58 &  8\\
J1313$+$0519 & 0.3--3 & 2.50$\pm$0.20 &  2.22$\pm$0.29 & 0.74 &  9\\
J1542$+$3100 & 0.3--4 & 2.41$\pm$0.18 &  2.17$\pm$0.28 & 1.30 &  9\\
J1559$+$3501 & 0.3--5 & 2.29$\pm$0.00 & 12.26$\pm$0.56 & 1.29 & 65\\
\enddata
\tablecomments{Col. (1): Abbreviated SDSS name. Col. (2): Energy range used in the fit.  Col. (3): Photon index of power law, where $N(E) \propto E^{-\Gamma_s}$. Col. (4): Normalization of the power law at 1 keV.  Col (5): Reduced $\chi^2$. Col. (6): Degrees of freedom.}
\label{tab:fit}
\end{deluxetable}

\end{document}